\documentclass[a4paper,12pt]{article}
\newcommand{\tline}{\setlength{\baselineskip}{0.84cm}}

\usepackage{epsfig}
\usepackage{lscape}
\begin{document}
\tline
\tline
\setcounter{page}{1}
\vspace{0.5cm}
\begin{flushright}
{\normalsize A4.08.211}
\end{flushright}
\begin{center}
{\large {\bf Novel Features in 2D and 3D Neutral,  Cationic and Anionic \\
Gold Clusters Au$_{5 \leq n \leq 9}^Z (Z = 0, \pm 1)$}}
\end{center}
\vspace{1cm}
\begin{center} 
F. REMACLE$^\dagger$\footnote[1]{Address for correspondence: FAX: +32 (4) 366 3413; E-mail addresses: fremacle@ulg.ac.be and \hspace*{0.5cm} eugene.kryachko@ulg.ac.be.}$^,$\footnote[2]{Ma\^itre de Recherche, FNRS (Belgium).} and E. S. KRYACHKO$^{\dagger \ddagger 1}$
\end{center}
\vspace{1cm}
{\normalsize \centerline{${}^\dagger$Departement of Chemistry, Bat. B6c, University of Liege}}
{\normalsize \centerline{Sart-Tilman, B-4000 Liege 1, Belgium}}
{\normalsize \centerline{and}}
{\normalsize \centerline{${}^\ddagger$Bogoliubov Institute for Theoretical Physics, Kiev, 03143 Ukraine}}
\vspace{2cm}
\begin{center}
{\bf Abstract}
\end{center}

Novel low-energy structures are found on the potential energy surfaces of the neutral, cationic and anionic gold clusters Au$_{5 \leq n \leq 8}^Z (Z = 0, \pm 1)$ and on the neutral potential energy surface of Au$_9$. These structures provide new insights on the 2D $\Rightarrow$ 3D transition in small neutral and charged gold clusters. It is demonstrated that the size threshold for the 2D - 3D coexistence is lower for cationic than neutral gold clusters: the 2D - 3D coexistence develops for Au$_5^+$ and Au$_7^+$ on the cationic potential energy surfaces while only for Au$_9$ on the neutral. Two metastable long-lived dianions of gold clusters are also reported. 
\newpage

\vspace{2cm}
\centerline{\bf {I. Introduction}}
\vspace{0.25cm}

Finite gold clusters are very common building blocks for nanostructured materials, electronic devices and novel nanocatalytic systems and as such, they have triggered numerous experimental and theoretical studies (Refs. [1, 2] and references therein). Some of them have shown that neutral and charged gold clusters favor 2D structures to unusually large sizes [3, 4] due very strong relativistic effects [5]. The motivation of the present work is to investigate in details the 2D - 3D coexistence for neutral clusters of gold and to extend it on the positively and negatively charged ones. Before discussing our own results and the new structures that we identify, we start this Introduction with a brief review of previous works on small gold clusters.

In one of the first ab initio calculations of small neutral and charged clusters of gold Au$_{4, 5}$, based on the size-consistent, SCF-based modified coupled pair functional (MCPF) method in conjunction with the Hay-Wadt 19-($5s^25p^65d^{10}6s$) valence electron relativistic effective core potential (RECP), Bauschlicher et al. [6] predicted the most stable Au$_5$ cluster to be a planar W-shape of {\it C}$_{2v}$ symmetry (see Figure 1). This geometry was later confirmed by Bravo-P\'erez et al. [7] within the second- and fourth-order perturbation M$\o$ller-Plesset (MP2 and MP4) method with the same Hay-Wadt RECP and by Gr\"onbeck and Andreoni [8] who used a variety of density functional (DF) methods. The 3D Au$_5$ clusters were reported to be higher in energy. Two low-energy trigonal bipyramid structures were found 0.42 and 0.95 eV above the W-shape in Ref. [7] (0.79 eV in Ref. [9] within the LDA PW91 method). The LSDA method with the Perdew-Zunger parametrization of the Ceperley-Alder data for the electron gas [8] also favors the trigonal pyramid as the most stable 3D structure of Au$_5$, though 0.72 eV higher than the W-form. The square pyramid is found 0.87 eV above the W-structure [8]. However, the BLYP density functional with the same RECP slightly reverses this order: the square pyramid lies 1.00 eV above the W-form while the trigonal pyramid is somewhat higher at 1.03 eV [8]. Using the BP86 DF method together with the Stuttgart 19e-RECP, Bona\v{c}i\'c-Kouteck\'y et al. [10] found that the C$_{2v}$ trigonal bipyramid is 0.930 eV higher than the W-structure. For anionic five-gold clusters, the square pyramid becomes more stable compared to the trigonal one (0.88 vs. 1.14 eV for LSDA and 1.05 vs. 1.25 eV for BLYP) [8].    

The global minimum of the six-gold cluster possesses a planar, 
D$_{3h}$-symmetry triangular structure [7]. The pentagonal pyramid is energetically higher by 0.44 eV [7]. Due to this smaller 2D - 3D energy difference compared to that predicted for five-gold clusters, it was then suggested that ``probably at $n = 7$ a 3D form could be the most stable or at least in degeneracy with a planar stable isomer" (Ref. [7b], p. 663). However, as reported in Ref. [10], the pentagonal pyramid lies 0.914 eV higher than the triangular cluster. In contrast, using the generalized gradient approximation (GGA), the pentagonal bipyramid is found to be the lowest energy structure, 0.16 eV below the planar hexagonal structure [9].   

H\"akkinen and Landman [3, 4] theoretically predicted that the 2D $\Rightarrow$ 3D transition occurs for Au$_n$ and Au$_n^-$ when $n \geq 8$ and $n \geq 7$, respectively (see also Ref. [11] for a study of the pyramidal Au$_7$ cluster with $D_{5h}$ symmetry and Ref. [12] for recent studies of the adsorption of methanol on gold clusters). It has recently been shown by Bona\v{c}i\'c-Kouteck\'y et al. [10] that 3D structures appear to be higher the 2D ones for $n = 7 - 10$ by 0.528 eV ($n = 7$), 0.530 eV ($n = 8$; for opposite results for $n = 7$ and 8 see the recent work [9]), 0.224 eV ($n = 9$), and 0.285 eV ($n = 10$). Interestingly, the differences in the binding energies per atom between the 2D ground-state and the lowest-energy 3D structures calculated in this work [10] fall from 0.19 eV for $n = 5$ and 0.15 eV for $n = 6$ to 0.02 - 0.03 eV for $n = 7 - 9$. A simple structural rule has been proposed in this study [10]: the ground-state structures of Au$_{4 \leq n \leq 7}$ can be derived by adding a single Au atom to the lowest-energy structure Au$_{n - 1}$ and forming an additional triangular subunit. This rule has been suggested to be a manifestation of the strong $s - d$ hybridization and of the preference for the directional $d$-character bonds over the isotropic $s$-ones. However, this structural motif disappears for Au$_8$ where the square is capped by four atoms of gold on each side [10]. The 2D - 3D boundary for the neutrals was recently further extended to at least $n = 13$ [3]. The recent gas-phase mobility experiments by Furche et al. [13] have demonstrated that planar anionic gold clusters are more stable for sizes as large as $n = 9 - 11$. On the other hand, similar experiments on the cationic gold clusters, supported by the GGA DF computations with the S-VWN+Becke-Perdew (BP86) parametrization together with the Stuttgart RECP, showed that Au$_n^+$ are 2D for $n = 3 - 7$ and 3D for $n = 8 - 13$ [14]. 

Understanding the 2D - 3D transition in small neutral and charged gold clusters is the main theme of the present work. The paper is organized as follows. Section II is devoted to a brief outline of the computational approach that is based on three  commonly used relativistic effective core potentials. In Sections III - VI we provide a systematic analysis of the potential energy surfaces (PESs) of Au$_{5 \leq n \leq 8}^Z (Z = 0, \pm 1)$ and identify therein a variety of novel structures that bring new insights on the 2D $\Rightarrow$ 3D transition. In Section VII we demonstrate the 2D - 3D coexistence on the PES of neutral Au$_9$. To conclude, the emerging trends of the 2D-3D transition as a function of the size, charge and odd or even number of electrons of the small gold clusters studied here are discussed in Section VIII. 
\vspace{1cm}

\centerline{\bf {II. Computational Methodology}}
\vspace{0.25cm}

All computations of gold clusters reported in the present work were carried out with the hybrid density functional B3LYP potential in conjunction with a variety of energy-consistent 19-($5s^25p^65d^{10}6s$) valence electron RECPs including the Los Alamos double-zeta type RECP LANL2DZ ($\equiv$ A) [15], the RECP  developed by Ermler, Christiansen and co-workers ($\equiv$ B) with the primitive basis set $(5s5p4d)$ [16], and the RECP of the Stuttgart group ($\equiv$ C) with the basis set $(8s7p6d)/[6s5p3d]$ [17] (see also Refs. [6, 7, 18] for the description of these RECPs). The reliability of the chosen RECPs has been assessed in Refs. [7, 10, 13, 14, 18-22] by comparing the computed results for small gold clusters (dimer, trimer and their charged species) with the experimental data. The GAUSSIAN 03 package of quantum chemical programs [23] was used throughout. All geometrical optimizations were conducted with the keyword ``tight". 

The B3LYP DF has recently been used for the gold clusters [20, 24-26] along with the others DFs: BP86 [10, 14], BLYP [8], PW91 [19], BPW91 [27], BPW [18]. Compared to the MP2 level, B3LYP was found [20] to give very similar geometries for clusters containing gold atoms, i.e., the Au-Au distances are $\approx$ 4 \% longer than the MP2 ones, with the B3LYP ones being more accurate, so that the density functional approach might be more realistic for the gold clusters than MP2 [20]. On the other hand, Bona\v{c}i\'c-Kouteck\'y et al. [10] and Lee et al. [18] have recently pointed out that the DF methods determine relatively small $s - d$ energy gaps and underestimate the gap E$_{HL}$ between the highly occupied molecular orbital (HOMO) and the lowest unoccupied molecular orbital (LUMO). For example, the $s - d$ orbital energy gap between the $d^{10}s^1$ and $d^9s^2$ configurations of Au$^-$ is estimated in the present work as equal to 0.88 eV (B3LYP/A), 1.12 eV (B3LYP/B), 1.37 eV (B3LYP/C), 0.23 eV (BP86/A), 0.43 eV (BP86/B), and 0.62 eV (BP86/C). These values are smaller than the experimental magnitude of 1.34 eV [28] except for the third estimate demonstrating a good performance of the B3LYP/C computational level for this task. To further pursue such comparisons for the geometrical, energetic, and MO patterns of larger gold clusters and to strengthen the credibility of the present conclusions, the BP86/A and PW91/A methods are also used for some key structures studied in this work. It is the well-known fact that DF methods favor planar structures for gold clusters (see Refs. [10, 18] and references therein). Since the present work focuses on the 2D - 3D transitions in the neutral and charged gold clusters studied by means of these methods, we have to estimate the DFT/RECP error in energy within which given Au clusters are treated as almost energetically equivalent. 

A value of the DFT/RECP error in energy based on the BP86/C method was proposed in Ref. [10]. It is defined as the energy difference between 2D and 3D structures for Au$_9$ clusters and equal to $\approx$ 0.2 eV ($\approx$ 4 kcal$\cdot$mol$^{-1}$; on p. 3124 of Ref. [10] this error is equal to 0.224 eV or 5.16 kcal/mol). As reported above, such a magnitude is in agreement with the error found in estimating of the $s - d$ orbital energy gap at the B3LYP/B and B3LYP/C compared to the experimental data. We suggest to take this value as the error margin for the B3LYP/RECP methods used here to study the 2D - 3D transitions in neutral and charged gold clusters. We futhermore demonstrate in Section VII, devoted to the neutral Au$_9$ clusters, that the two criteria, viz., the energy difference between 2D and 3D structures as used in Ref. [10] and the $s - d$ orbital gaps lead indeed to the same value of 0.19 - 0.2 eV ($\approx$ 4.33 - 4.55 kcal/mol, see Table 2). 

The harmonic vibrational frequencies and corresponding zero-point vibrational energies (ZPVE) were calculated at all employed B3LYP/X levels (X = A, B, and C) in order to determine the topology of the stationary points. The electronic energies calculated by means of the B3LYP DF with the basis sets A, B, and C are hereby reported in the following order: Energy$_{B3LYP/A}$ (Energy$_{B3LYP/A}$ + ZPVE) [Energy$_{B3LYP/B}$ (Energy$_{B3LYP/B}$ + ZPVE); Energy$_{B3LYP/C}$ (Energy$_{B3LYP/C}$ + ZPVE)]. The BP86/A values are given in curly brackets whereas the PW91/A ones appear in the double curly 
brackets. Similar notations are used for the optimized geometrical parameters.

The vertical electron detachment energy, VDE, is calculated as the total energy difference between the anionic and neutral clusters, both taken in the anionic optimized geometry without the ZPVE. The adiabatic electron affinity, EA$_a$ and ionization energy, IE$_a$, are calculated as the total energy difference between the anionic and cationic clusters, respectively, and the parent neutral cluster in their optimized geometries.  
\vspace{1cm}

\centerline{\bf {III. Au$_5^Z$}}
\vspace{0.5cm}

The PES of the neutral gold cluster Au$_5$ includes four low-energy planar structures that are shown in Figure 1. In full agreement with all the earlier studies [4, 6-10, 19, 29], at all the employed computational levels, the global minimum is the nonpolar, trapezoidal W-shape cluster Au$_5^I$ of {\it C}$_{2v}$ symmetry in the electronic state ${}^2A_1$. Its geometry is compatible with the triangular motif rule. Au$_5^I$ has one 4-coordinated Au atom and two 3- and two 2-coordinated ones. The Au-Au distances in Au$_5^I$ are partitioned into four classes (see Table 1): the largest ones, 2.83 - 2.87, {\AA} separate 4- and 3-coordinated atoms; those between 3-coordinated atoms are 2.74 - 2.78 {\AA}, and those between 2- and 3-coordinated atoms 2.72 - 2.75 {\AA}. The 2- and 4-coordinated atoms are connected by the shortest bonds with the bond lengths of 2.68 - 2.70 {\AA}. The spin-up and spin-down gaps, E$_{HL \uparrow}$ and E$_{HL \downarrow}$, are equal to 1.58 (1.57) and 1.01 (1.02) eV, respectively at the BP86/A (PW91/A) computational levels (cf. with E$_{HL \uparrow}$ = 1.51 eV in Ref. [9] and 1.37 eV in Ref. [29]). The B3LYP DF method estimates them at about 2.5 and 2.2 eV, respectively (Table 2). 

The next structure, by increasing energy, is the X-shape nonpolar cluster Au$_5^{II}$ in the ${}^2B_{2u}$ state of {\it D}$_{2h}$ symmetry, lying 7.3 - 8.2 kcal$\cdot$mol$^{-1}$ above the ground W-shape state (see Table 2). These energies fall within the earlier reported range: 7.84 - 14.53 kcal$\cdot$mol$^{-1}$ [8] and 9.75 kcal$\cdot$mol$^{-1}$ [10]. Compared to Au$_5^{I}$, Au$_5^{II}$ possesses a narrower spin-up HOMO-LUMO gap (0.77 eV for the BP86/A; 1.4 - 1.5 eV for the B3LYP) and a slightly wider one for the spin-down electrons (1.31 eV for the BP86/A; 2.7 - 3.0 eV for the B3LYP; see Table 2). 

Two higher-energy clusters, Au$_5^{III}$ and Au$_5^{IV}$ (Figure 1), are reported in the present work for the first time. As follows from Table 2, they fill the energetic gap between the 2D clusters Au$_5^{I}$ and Au$_5^{II}$ on one hand and, on the other, the 3D pyramid which is ca. 21.4 kcal$\cdot$mol$^{-1}$ above Au$_5^{I}$ [10]. Both novel structures are less compact, less coordinated, polar (1.2 D and $\approx$ 4.2 D, respectively) and characterized by rather large polarizabilities (larger by a factor of 1.5 - 1.6 compared to that of Au$_5^I$). Their HOMO-LUMO gaps are given in Table 2. One of them, Au$_5^{III}$, is rather important since it gives rise to the most stable anion Au$_5^{III-}$ (Figure 1). Au$_5^{III-}$ is a novel structure too (see Tables 1 and 2) and is 5.8 - 7.2 kcal$\cdot$mol$^{-1}$ lower in energy than the well-known cluster Au$_5^{I-}$ [4, 8, 13, 30], which is well beyond the estimated error of ca. 4 kcal$\cdot$mol$^{-1}$ of the DF/RECP methods. 

The lower energy of Au$_5^{III-}$ is readily explained by its larger interatomic distances r(Au$_2$-Au$_5$) = 6.807 [6.803; 6.889] {\AA} and r(Au$_2$-Au$_3$) = r(Au$_3$-Au$_5$) = 5.300 [5.250; 5.296] {\AA}, which reduces the mutual Coulomb repulsion between the partial excess electron charges that reside mainly on these atoms (the Mulliken charges are equal to -0.37 [-0.49; -0.55], -0.23 [-0.28; -0.23], and -0.37 [-0.49; -0.55] on Au$_2$, Au$_3$, and Au$_5$, respectively). On the other hand, in Au$_5^{I-}$, the excess electron spreads over Au$_2$ (-0.24 [-0.19; -0.12]), Au$_3$ (-0.31 [-0.40; -0.26]), Au$_4$ (-0.31 [-0.40; -0.26]), and Au$_5$ (-0.24 [-0.19; -0.12]), where r(Au$_2$-Au$_3$), r(Au$_2$-Au$_5$), r(Au$_4$-Au$_5$) are only within 2.72 - 2.78 {\AA}. In Au$_5^{III-}$, the distance r(Au$_1$-Au$_4$) = 2.7 - 2.9 {\AA} is smaller by factor of two than the van der Waals radius of Au implying that Au$_1$ forms the Au-Au bond with Au$_4$ and hence, Au$_5^{III-}$ is distinct from the V-structure shown in Figure 4 of Ref. [18] (see also note [31]).

These two low-energy anions Au$_5^{I-}$ and Au$_5^{III-}$ also differ by their MO patterns and therefore have different catalytic reactivities. Within the B3LYP method, the Au$_5^{I-}$ HOMO-1 - HOMO and HOMO-LUMO gaps are equal to 0.94 [1.01; 1.00] eV (see note [32] and Table 2) and 1.21 [0.93; 1.09] eV, respectively. In Au$_5^{III-}$ the B3LYP HOMO-1 - HOMO gap is substantially narrower, viz., 0.27 [0.23; 0.30] eV while the HOMO-LUMO one is wider and equal to 2.88 [2.39; 2.41] eV. The EA$_a$s and VDEs of these anionic clusters are the following: EA$_a$(Au$_5^{I-}$) = 3.18 [3.02; 3.07] eV; EA$_a$(Au$_5^{III-}$) = 4.09 [3.96; 3.99] eV; VDE(Au$_5^{I-}$) = 3.29 [3.09; 3.16] eV and VDE(Au$_5^{III-}$) = 4.20 [4.09; 4.06] eV. The EA$_a$ and VDE of Au$_5^{I-}$ are in agreement with the experimental and earlier theoretical data: EA$_a^{expt}$(Au$_5$) = 2.992 $\pm$ 0.050 eV [33], 2.93 $\pm$ 0.10 eV [34], 3.06(3) eV [30]; EA$_a^{theor}$(Au$_5$) = 3.33 eV [6], 3.33 eV (LSDA [8]), 3.04 eV (BLYP [8]), 3.06 eV [27]; VDE$^{expt}$(Au$_5^-$) = 2.98 $\pm$ 0.2 eV [33], 3.12 $\pm$ 0.05 eV [34], 3.09(3) eV [30]; VDE$^{theor}$(Au$_5^-$) = 3.35 eV (LSDA [8]), 3.08 eV (BLYP [8]), 3.2 eV (obtained with the BP86 and the RECP C augmented by additional polarization and diffuse functions optimized for the neutral Au$_3$ cluster [13]), 3.06 eV [18], and 3.09 eV [30]. Obviously, such comparison is fairly satisfactory. Let us recall that the neutral parent of Au$_5^{I-}$ is the ground-state cluster Au$_5^{I}$ whereas the neutral parent of Au$_5^{III-}$ is the higher-energy isomer Au$_5^{III}$. On the neutral PES, the transition state between Au$_5^{I}$ and Au$_5^{III}$ involves the breaking of two Au-Au bonds and its barrier is estimated to be as large as 15 kcal$\cdot$mol$^{-1}$ (see Table 2). Since a similar transition barrier exists on the anionic PES, Au$_5^{I-}$ is metastable and that may be why it is this isomer that has been observed experimentally. It would be interesting to detect experimentally the actual ground-state anion Au$_5^{III-}$ reported here for the first time (see note [35]).

We also demonstrate that two electrons can be simultaneously attached to Au$_5^{III}$ to form the dianion Au$_5^{III 2-}$ shown in Figure 1. Its Au-Au bond lengths are collected in Table 1. This dianion is stable relative to Au$_5^{III}$ by 65.6 kcal$\cdot$mol$^{-1}$ (65.8 kcal$\cdot$mol$^{-1}$ after ZPVE) although it still remains unstable with respect to the corresponding anion Au$_5^{III-}$ by 28.8 kcal$\cdot$mol$^{-1}$ (28.6 kcal$\cdot$mol$^{-1}$ with ZPVE). It is responsible for the second adiabatic electron affinity (see note [32]). Such dianion may therefore exist only in solvent or condensed phase (see Ref. [36] for current review) rather than in the gas phase implying that there is no contradiction with the statement that the smallest stable dianion is Au$_{12}^{2-}$ [37]. 

Ionization of all low-energy clusters Au$_5^{I-IV}$ yields the same well-known planar X-shape cation Au$_5^{I+}$ of {\it D}$_{2h}$ symmetry [14] displayed in Figure 1. Its bond lengths, evaluated at the B3LYP/X (X = A, B, C) computational levels, are slightly larger, by 0.08 {\AA}, compared to those obtained by Gilb et al. [14] using the DF BP$86$ 
in conjunction with the RECP C augmented by additional polarization and diffuse functions optimized for the neutral Au$_3$ cluster (see Table 1). The enthalpy $\Delta$H$_f$ of formation of Au$_5^{I+}$ from the ground-state pentamer cluster Au$_5^I$ is estimated being equal to 7.16 [7.13; 7.09] eV. It is consistent with the B3LYP adiabatic ionization energy IE$_a^{B3LYP}$(Au$_5$) = 7.16 [7.13; 7.09] eV. These estimates are close to the experimental data on IE$^{expt}$(Au$_5$) = 8.00 (upper limit) [21, 22]; 7.61 $\pm$ 0.20 eV [40] and earlier theoretical data: IE$_v^{theor}$(Au$_5$) = 7.46 [6], 7.5 [18], 7.68 [21, 22], and IE$_a^{theor}$(Au$_5$) = 7.56 (7.72) [41] and 7.78 (7.83) eV [9]. While the ionization pathways from Au$_5^{I}$ and Au$_5^{II}$ to Au$_5^{I+}$ are direct although the former is related to the Au$_2$-Au$_5$ bond breaking, the pathway Au$_5^{III} \Rightarrow $ Au$_5^{I+}$ goes via a transition state lying $\approx$ 15 kcal$\cdot$mol$^{-1}$ higher than the product and is similar to the one that connects Au$_5^{I}$ and Au$_5^{III}$. 

The cationic PES of the five-gold clusters exhibits the first indication of the 2D - 3D coexistence at the B3LYP/A and B3LYP/B computational levels (see also Refs. [12b, 41]). Together with the planar Au$_5^{I+}$, it contains the 3D rotamer Au$_5^{I+ \prime}$ (see Figure 1) characterized by dihedral angle $\angle$Au$_2$Au$_3$Au$_4$Au$_5 = 82.4^o$ and bond lengths r(Au$_1$-Au$_2$) = 2.768 and 2.629 {\AA} and r(Au$_2$-Au$_3$) = 2.645 and 2.629 {\AA} for the B3LYP/A and B3LYP/B methods, respectively. These bond lengths are very close to those in Au$_5^{I+}$. Au$_5^{I+ \prime}$ is almost isoenergetic to Au$_5^{I+}$. It is 0.3 kcal$\cdot$mol$^{-1}$ higher at the B3LYP/A and 0.1 kcal$\cdot$mol$^{-1}$ at the B3LYP/B after ZPVE (cf. 2.08 kcal$\cdot$mol$^{-1}$ in Ref. [41]). The B3LYP/C level predicts this rotamer to be exactly energetically equivalent to Au$_5^{I+}$, though with a smaller dihedral angle, $\angle$Au$_2$Au$_3$Au$_4$Au$_5 = 18.0^o$. The difference in entropy at room temperature between the 2D and 3D cation of Au$_5^{I}$ varies between 0.9 - 1.0 kcal/mol, depending on a level of computations, that together with the corresponding enthalpy difference of $\approx$ 0.01 eV gives a minor contribution to that of the B3LYP Gibbs free energies of formation of Au$_5^{I+}$ and Au$_5^{I+ \prime}$. 
\vspace{1cm}

\centerline{\bf {IV. Au$_6^Z$}}
\vspace{0.25cm}

The global-minimum energy cluster Au$_6^{I}$ is well separated from the other isomers, including the 3D ones in particular (see Figure 2 and Table 3). Structurally, Au$_6^{I}$ retains the triangular motif and is obtained from Au$_5^{I}$ by adding another triangle to the Au$_2$-Au$_5$ bond of Au$_5^{I}$ (Figure 1). It also retains the parent nonpolarity. Note that the BP86/A and B3LYP/B predict almost equivalent geometries of Au$_6^{I}$ (see Table 3 for details). The addition of one gold atom to Au$_5^{I}$ elongates its Au$_2$-Au$_5$ bond of by $\approx$ 0.13 - 0.16 {\AA} (as the bond between 4-coordinated atoms) and shortens the bonds Au$_2$-Au$_3$ and Au$_4$-Au$_5$ by 0.04 - 0.05 {\AA}. Energetically, the binding energy of Au$_6^{I}$ relative to the asymptote Au$_5^{I}$ + Au$_1$ is equal to 60.4 (60.0 after ZPVE) [61.3 (60.9); 59.4 (59.0)] kcal$\cdot$mol$^{-1}$ (cf. 50.3 kcal$\cdot$mol$^{-1}$ in Ref. [9] and 63.9 (54.2) kcal$\cdot$mol$^{-1}$ in Ref. [29]). 

As expected, since it is a closed-shell cluster characterized by a larger binding energy of the paired valence electrons compared to the single valence one, Au$_6^{I}$ is more inert than Au$_5^{I}$. This is confirmed by larger E$_{HL}$ of 2.23 eV (BP86/A) (B3LYP: 3.48 [3.35; 3.48] eV). The BP86/A magnitude of the E$_{HL}$ is very close to the experimental value of $\approx$ 2.5 eV reported by Taylor et al. [42, 33] (compare with the earlier theoretical data: 2.05 eV [4, 6], 2.06 eV [9] and 1.74 eV [29]; see also the recent work [43]). The IR strongest Au-Au stretching mode (doublet; 6 [5; 5] km$\cdot$mol$^{-1}$) is predicted at 167 [174; 167] cm$^{-1}$. 

Under ionization Au$_6^{I}$ retains its planar triangular structural form [14]. Figure 2 displays the cation Au$_6^{I+}$ whose geometry is compared to that of Au$_6^{I}$ in Table 3. Inspecting Table 3, it is worth mentioning that the B3LYP/B bond lengths of Au$_6^{I+}$ practically resemble those calculated in Ref. [14], using the DF BP86 and the RECP C augmented by additional polarization and diffuse functions optimized for the neutral Au$_3$ cluster, within the margin of 0.02 - 0.06 {\AA}. The adiabatic ionization energy IE$_a^{B3LYP}$(Au$_6^I$) = 8.46 [8.28; 8.33] eV is about 1 eV larger than IE$_a^{B3LYP}$(Au$_5^I$) and satisfactorily agrees with recent experimental data, IE$^{expt}$(Au$_6$) = 8.80 eV (upper limit) [35, 36], and IE$^{theor}$(Au$_6$) = 7.60 [11], 8.55 [9], 8.17 [18], and 8.37 eV [21, 22]. 

We do not find any new clusters on the low-energy section of the PES of Au$_6$. Notice that, as mentioned in Introduction, the pentagonal pyramid lies 0.44 eV higher Au$_6^I$ (MP2/Hay-Wadt RECP) [7]. The energy difference increases to 0.65 eV [18] while using the MP2 method in conjunction with the 19-electron Stuttgart-type RECP C$^\prime$ defined in Ref. [18] and to 0.89 eV at the BPW/B computational 
level. However, new ones are identified on the respective cationic and anionic PESs. There are two low-energy structures on the anionic PES (see Figure 2 and Table 3). One of them is the known planar hexagonal structure Au$_6^{I-}$ [4, 18, 33, 42, 44] structurally resembling Au$_6^{I}$. Its VDE(Au$_6^{I-}$) = 2.27 [2.18; 2.20] eV and EA$_a$(Au$_6^{I}$) = 2.16 [2.09; 2.09] eV (BP86/A: 2.43 eV satisfactorily agree with the experimental data: VDE$^{expt}$(Au$_6^-$) = 2.29 eV [4], 2.0 eV [33], 2.13(2) eV [30] (VDE$^{theor}$(Au$_6^-$) = 2.35 eV (obtained with the BP86 and the RECP C augmented by additional polarization and diffuse functions optimized for the neutral Au$_3$ cluster [13]), 2.24 eV [18], 2.29 eV [30]) and EA$_a^{expt}$(Au$_6$) = 2.0510 $\pm$ 0.0020 [45]; 2.016 $\pm$ 0.050 [33]; 2.06(2) eV [30] (EA$_a^{theor}$(Au$_6$) = 2.22 eV [30]). The gap E$_{HL \uparrow}$ of Au$_6^{I-}$ significantly decreases to 0.4 eV (BP86/A) (B3LYP gives 1.11 [1.15; 1.10] eV) compared to the neutral parent whereas the E$_{HL \downarrow}$ holds approximately at the same value of 2.09 eV (BP86/A) (B3LYP: 3.30 [3.27; 3.27] eV). The electronic energy and enthalpy of the reaction Au$_5^{I-}$ + Au $\Rightarrow$ Au$_6^{I-}$ are equal to 36.7 (36.4 after ZPVE) and 42.6 kcakcal$\cdot$mol$^{-1}$ (B3LYP/A). These values are within the experimentally measured margin $< 62.5 \pm 9.2$ kcal$\cdot$mol$^{-1}$ [46]. 

The other anion Au$_6^{II-}$ (Figure 2 and Table 3) is directly related to the most stable cluster Au$_5^{III-}$ on the anionic PES of Au$_5$. Actually, adding a single Au atom to the Au$_3$ atom of Au$_5^{III-}$ in Figure 1 yields Au$_6^{II-}$ with a gain of energy of 26.9 (26.7) kcal$\cdot$mol$^{-1}$ (B3LYP/A). The anion Au$_6^{II-}$ is however less stable, viz., by 2.5 (2.5) [5.8 (5.7); 4.4 (4.3)] kcal$\cdot$mol$^{-1}$ (BP86/A: 6.7 kcal$\cdot$mol$^{-1}$) than Au$_6^{I-}$. Its HOMO-LUMO gaps amount to 1.91 eV (spin-up, BP86/A; B3LYP: 2.80 [2.82; 2.84] eV) and 0.86 eV (spin-down, BP86/A; B3LYP: 1.84 [1.86; 1.74] eV). Notice finally that the reaction pathway Au$_6^{I} + e^- \Rightarrow$ Au$_6^{II-}$ is indirect and occurs through a barrier.

There are three low-energy clusters Au$_6^{I+}$, Au$_6^{III+}$, and Au$_6^{IV+}$ on the cationic PES shown in Figure 2 (see also Table 3 for their geometrical parameters). In fact, they are all cationic isomers of the neutral Au$_6^{I}$ resulting from the barrierless (Au$_6^{I+}$) or barrier-type ionization pathways. The former two are the known planar clusters [14] although the used B3LYP computational level slightly favors Au$_6^{III+}$ over Au$_6^{I+}$ (by 0.9 (1.0) [0.3 (0.3); 1.4 (1.4) kcal$\cdot$mol$^{-1}$]; an opposite, as though negligible trend, of -0.03 kcal$\cdot$mol$^{-1}$, is predicted in Ref. [14] for the DF BP86 with the RECP C augmented by additional polarization and diffuse functions optimized for the neutral Au$_3$ cluster). In terms of the Gibbs free energy, the present order holds (with the corresponding difference of about 1 kcal/mol) at room temperature. The third one, Au$_6^{IV+}$, is 3D and somewhat less stable (by 5 - 7 kcal$\cdot$mol$^{-1}$) than Au$_6^{I+}$. Since this energy difference is close to the estimated energy error of the DF methods, this indicates that the 2D - 3D demarcation line appears at lower energies for the positively charged even-sized gold clusters than for their neutral even-sized parents. 
\vspace{1cm}

\centerline{\bf {V. Au$_7^Z$}}
\vspace{0.25cm}

Adding a single gold atom to Au$_6^{I}$ and holding the planar triangular motif rule, one obtains the known global-minimum energy structure Au$_7^{I}$ in the electronic ${}^2A^\prime$ state of C$_s$ symmetry characterized by a stabilization energy equal to 24.6 (24.4) [26.8 (26.6); 24.6 (24.4] kcal$\cdot$mol$^{-1}$. This structure is shown in Figure 3 and its optimized geometrical parameters are gathered in Table 4 including those computed within the BP86/A method. Au$_7^{I}$ is an open-shell cluster, slightly polar (0.4 [0.3; 0.3] D), having the following spin-up and spin-down HOMO-LUMO gaps: 1.32 and 1.04 eV (BP86/A), respectively (the B3LYP: 2.24 [2.19; 2.20] and 2.25 [2.20; 2.27] eV). 

Our computations show that seven is the smallest number of gold atoms leading to a cluster whose neutral PES contains a low-lying 3D structure. This is the novel 3D Au$_7^{II}$ cluster that is higher in energy than Au$_7^{I}$ by about 5 - 6 kcal$\cdot$mol$^{-1}$, viz., 5.6 (5.5) [6.5 (6.5); 5.7 (5.4)] kcal$\cdot$mol$^{-1}$ (the BP86/A gives 5.0 (5.0) kcal$\cdot$mol$^{-1}$) [47]. Figure 3 shows that the three dimensionality arises from capping one of edge triangles of Au$_6^{I}$ with a gold atom (called `edge-capped pyramid' by analogy with Ref. [3]). The 3D structure Au$_7^{II}$ has twice the polarity (0.8 [0.9; 0.9] D) of Au$_7^{I}$. Its gap E$_{HL \uparrow}$, equal to 0.92 eV (BP86/A) (B3LYP predicts a wider gap of 1.83 [1.74; 1.78] eV), is responsible for its higher reactivity. The spin-down gap E$_{HL \downarrow}$ is larger, namely, 1.79 eV at the BP86/A level (cf. the B3LYP estimates it as equal to 3.04 [3.11; 3.10] eV). 

The 2D - 3D coexistence between the two structures, Au$_7^{I}$ and Au$_7^{II}$, appears more clearly on the cationic PES shown in Figure 3 (see also Table 4). Upon ionization, the 2D cluster Au$_7^{I}$ becomes the 3D cation Au$_7^{II+}$ via the 2D barrier Au$_7^{I+ts}$, which is structurally similar to its parent neutral. Au$_7^{II+}$ is similar to the 3D neutral Au$_7^{II}$ (Figure 3). The transition-state linker Au$_7^{I+ts}$ lies 4.8 kcal$\cdot$mol$^{-1}$ higher (B3LYP/A). This is one remarkable facet of the cation Au$_7^{II+}$. The other is that ionization of Au$_7^{II}$ directly yields Au$_7^{II+}$. Such a 2D - 3D geometry change upon ionization is reported in this work for the first time. The Au$_7^{I}$ $\Rightarrow$ Au$_7^{I+ts}$ $\Rightarrow$ Au$_7^{II+}$ pathway determines the ionization energy IE$_a^{B3LYP}$(Au$_7^I$) = 6.99 [6.94; 6.91] eV which is smaller than IE$_v^{expt}$(Au$_7$) = 7.80 eV [35, 36] and rather close to the theoretical value IE$_v^{theor}$(Au$_7$) = 7.18 given in Ref. [18]. 

Upon electron attachment, the low-energy 3D cluster Au$_7^{II}$ found in this work exhibits a 3D to 2D conversion that leads to the novel planar structure Au$_7^{II-}$. It is shown in Figure 3 (see also Table 4 for its geometry) and characterized by rather large VDE(Au$_7^{II-}$) = 4.50 [4.34; 4.42] eV. The latter clearly indicates that a 3D structure for the anion is less stable than the less strained 2D one. The HOMO-LUMO gap of Au$_7^{II-}$ is equal to 1.72 eV (BP86/A) (the B3LYP: 2.85 [2.77; 2.79] eV). However, Au$_7^{II-}$ does not occupy the global minimum on the anionic PES of Au$_7$, rather it lies above the well-known ground-state tricapped square cluster Au$_7^{I-}$ [3, 30] by 4.6 (4.5) [7.9 (7.7); 6.0 (5.9)] kcal$\cdot$mol$^{-1}$ (the BP86/A estimate is 10.5 (10.3) kcal$\cdot$mol$^{-1}$). Due to a large entropy effect, the difference in the B3LYP Gibbs free energies between Au$_7^{II-}$ and Au$_7^{I-}$, evaluated at T = 298 K, is rather small and equal to 1.9 [4.7; 2.2] kcal$\cdot$mol$^{-1}$, reflecting the floppy nature of the bottom of the anionic PES. 

The anionic PES exhibits two energetically almost degenerate 2D structures: Au$_7^{I-}$ and Au$_7^{III-}$ shown in Figure 3 (see Table 4 and note [48]). The latter is a novel structure obtained by adding a single gold atom to one of the edge atoms of Au$_6^{I-}$. Actually, it lies 2.5 (2.5) [1.7 (1.7); 2.4 (2.4)] kcal$\cdot$mol$^{-1}$ (the BP86/A: 0.9 (0.9) kcal$\cdot$mol$^{-1}$) below Au$_7^{I-}$. Its VDE(Au$_7^{III-}$) = 3.87 [3.74; 3.74] eV and its HOMO-LUMO gap is equal to 1.35 (BP86/A) (whereas the B3LYP gives 2.17 [2.17; 2.12] eV). 

The parent neutral cluster of Au$_7^{III-}$ is the honeycomb Au$_7^{III}$ in the state ${}^2B_{2g}$ of $D_{2h}$ symmetry [10]. On the neutral PES, Au$_7^{III}$ lies however 9.0 (8.8) [7.7 (7.5); 9.0 (8.8)] kcal$\cdot$mol$^{-1}$ above the ground state (the BP86/A value is 8.1 (7.9) kcal$\cdot$mol$^{-1}$; cf. with 7.0 kcal$\cdot$mol$^{-1}$ in Ref. [10]; see also note [49]). In contrast to Au$_7^{I}$, Au$_7^{III}$ is nonpolar and its spin-split HOMO-LUMO gaps are equal to 0.18 and 1.73 eV (BP86/A) (cf. with the B3LYP predictions: 1.01 [0.95; 0.97] and 2.76 [2.76; 2.78] eV and the value of 0.87 eV reported in Ref. [29]). 

Compared to Au$_7^{II-}$, Au$_7^{I-}$ is more reactive due to a smaller E$_{HL}$ gap = 1.18 eV (BP86/A) (the B3LYP: 1.95 [1.94; 1.89] eV) and is characterized by an EA$_a$ = 3.40 (3.40) [3.29 (3.29); 3.28 (3.28)] eV (3.62 eV for the BP86/A) and a VDE = 3.50 [3.39; 3.38] eV. Both these computed values are fairly close to EA$_a^{expt}$(Au$_7$) = 3.415 $\pm$ 0.050 [33] and 3.40(3) eV [30] and VDE$^{expt}$(Au$_7^-$) = 3.46(2) eV [3, 30] and 3.38 eV [33], respectively. They also fairly agree with previous theoretical data: EA$_a^{theor}$(Au$_7$) =  3.43 eV [30] and VDE$^{theor}$(Au$_7^-$) = 3.55 eV (obtained with the BP86 and the RECP C augmented by additional polarization and diffuse functions optimized for the neutral Au$_3$ cluster [13]), 3.43 eV [18], and 3.46 eV [30].     

Let us return to the cationic PES of Au$_7$ sketched in Figure 3. Two novel different structures are found on its lowest-energy portion. One, as described above, is the cation Au$_7^{I+}$ and the other is a 3D honeycomb-type Au$_7^{III+}$ (see Table 3 and note [50]). The energy difference between Au$_7^{III+}$ and Au$_7^{I+}$ is 0.1 (0.2) [2.9 (2.9); -1.9 (-2.0)] kcal$\cdot$mol$^{-1}$ [51] meaning that the computational level B3LYP/C slightly favors the cation Au$_7^{III+}$. As shown in Figure 3, this cation Au$_7^{III+}$ is not perfectly planar at the B3LYP/A and B3LYP/B computational levels which predict the dihedral angle $\angle$Au$_2$Au$_3$Au$_7$Au$_6 = 31.6^o$ and $31.5^o$, respectively (B3LYP/C gives only 0.2$^o$ and BP86/B does 27.5$^o$). In summary, it is found for the first time that there may be no 2D clusters in the low-energy section (likely up to the energies of ca. 15 kcal$\cdot$mol$^{-1}$; see Ref. [10] for the corresponding neutrals) of the cationic PES of seven gold atoms. 

To end this Section, let us mention another neutral structure Au$_7^{IV}$ in the ${}^2A_1$ state of $C_{2v}$ symmetry (Figure 3 and Table 3) lying 8.5 (8.6) kcal$\cdot$mol$^{-1}$ (B3LYP/A) above Au$_7^{III}$ (cf. with 8.3 kcal$\cdot$mol$^{-1}$ reported in Ref. [10]). It possesses spin-up and spin-down HOMO-LUMO gaps of 1.52 and 1.50 eV, respectively.  The related transition-state double-rhombus structure Au$_7^{IVts}$ is also shown in this Figure 3 and determines a rather high barrier of $\approx$17.7 kcal$\cdot$mol$^{-1}$ governing the breaking of the Au$_2$-Au$_5$ bond.
\vspace{1cm}

\centerline{\bf {VI. Au$_8^Z$}}
\vspace{0.25cm}

It is known that the Au$_8$ cluster is the smallest catalytically active size of gold clusters [1e].

A binding of a single gold atom Au$_1$ to the Au$_4$-Au$_5$ bond of the Au$_7^I$ cluster leads to the planar and nonpolar tetracapped square cluster Au$_8^I$ (Figure 4) which is the global minimum on the neutral PES of eight atom gold clusters. Such a structure has recently been reported by Bona\v{c}i\'c-Kouteck\'y et al. [10, 21] as the true global minimum, assigned to the ${}^1A_{1g}$ state of $D_{4h}$ symmetry, in contrast to earlier findings. The binding energy of Au$_8^I$, taken relative to the infinitely separated Au$_7^I$ and Au$_1$, is equal to 56.0 (55.6) [56.8 (56.5); 54.6 (54.2)] eV. The E$_{HL}$ gap width amounts to 1.31 eV (BP86/A) (the B3LYP gives 2.77 [2.72; 2.67] eV). 

Structurally, Au$_8^I$ is composed of two kinds of coordinated gold atoms: the four edged atoms are 2-coordinated while the other four (central square) are 4-coordinated. The Au-Au bond length between 2- and 4-coordinated atoms is about 0.15 - 0.17 {\AA} (see Table 4) shorter than between each pair of 4-coordinated atoms. Each pair of opposite atoms belonging to the central square are unbounded, forming thus a sort of hole since the corresponding distances of ca. 4.0 {\AA} exceed twice the van der Waals radius of gold (3.32 {\AA}). It is worth noticing that the formation of Au$_8^I$ from Au$_7^I$ + Au$_1$ requires to break an Au-Au bond of length $\approx$ 3.0 {\AA} ($< 3.32$ {\AA}) in the internal rhombus of Au$_7^I$. 

The two following low-energy structures, Au$_8^{II}$ and Au$_8^{III}$, have almost the same energy but differ by their 2D - 3D character (see Figure 4 and Table 2). They lie about 11 - 12 kcal$\cdot$mol$^{-1}$ above Au$_8^I$ : $\Delta$E(Au$_8^{II}$ - Au$_8^I$) = 12.5 (12.4) [10.8 (10.7); 11.0 (10.9)] kcal$\cdot$mol$^{-1}$ (and 9.7 (9.6) kcal$\cdot$mol$^{-1}$ for the BP86/A) and $\Delta$E(Au$_8^{III}$ - Au$_8^I$) = 12.7 (12.6) [13.9 (13.8); 12.8 (12.7)] kcal$\cdot$mol$^{-1}$ and are of the same polarity ($\approx$ 0.9 D). One of them, Au$_8^{II}$, is a planar $C_{2v}$ structure that possesses a sixfold coordinated atom at the center. The other, Au$_8^{III}$, is a novel 3D cluster formed by a Au$_2$ dimer bonded to the six-gold most stable cluster Au$_6^{I}$ along its symmetry axis. It possesses four fivefold coordinated atoms of gold. The difference between the two clusters is also manifested in the width of E$_{HL}$: while for the 3D cluster Au$_8^{III}$, it is closer to that of Au$_8^{I}$, viz., 1.86 eV (BP86/A) (B3LYP: 2.95 [2.85; 2.94] eV), its 2D counterpartner has a narrower one, i. e., 0.87 eV (BP86/A) (B3LYP: 1.76 [1.71; 1.72] eV). These two structures are slightly below (ca. 3 - 4 kcal$\cdot$mol$^{-1}$ depending on the computational level) a pair of less stable novel 3D clusters, Au$_8^{IV}$ and Au$_8^{V}$ shown in Figure 4. Au$_8^{IV}$ is formed by binding a gold dimer to Au$_6^{I}$ along its 3-atom bond whereas in Au$_8^{V}$ a gold trimer parallel is attached parallel to Au$_5^{I}$ (see Figure 1). 

While the 3D cluster Au$_8^{IV}$ is rather far from the global minimum of the neutral PES, on the cationic one, Au$_8^{IV+}$ competes with the 2D Au$_8^{I+}$ and Au$_8^{II+}$, and the three isomers lie at the very bottom of the PES (see Figure 4). More precisely, Au$_8^{IV+}$ is only 0.1 (0.1) [2.3 (2.3); 1.2 (1.2)] kcal$\cdot$mol$^{-1}$ above Au$_8^{I+}$ which in turn is only 1.7 (1.7) [0.6 (0.6); 0.4 (0.4)] kcal$\cdot$mol$^{-1}$ lower than Au$_8^{II+}$ [52]. This is a manifestation of the 2D $\Leftrightarrow$ 3D transition that occurs on the cationic PES of eight gold atoms. The optimized geometries and some other properties of these three cationic clusters are presented in Table 5. Notice that the B3LYP/B bond lengths of Au$_8^{II+}$ are in a good agreement (within the margin of 0.03 - 0.04 {\AA}) with the bond lengths obtained in Ref. [14] using the BP86 and the RECP C augmented by additional polarization and diffuse functions optimized for the neutral Au$_3$ cluster. The IE$_a$(Au$_8^{I}$) = 7.88 [7.71; 7.72] eV correlates with the experimental upper limit of 8.65 eV [35]. The IE$_a$(Au$_8^{II}$) = 7.41 [7.27; 7.26] eV. 

That Au$_8^{I-}$ is the most stable anionic cluster is a well established fact [13]. Its structure is displayed in Figure 4. Two other low-energy clusters: the planar Au$_8^{II-}$ which is 7.1 (7.0) [6.9 (6.8); 6.6 (6.6)] kcal$\cdot$mol$^{-1}$ higher than Au$_8^{I-}$ and the three-dimensional Au$_8^{III-}$ lying 21.1 (20.9) [23.4 (23.2); 21.0 (20.8)] kcal$\cdot$mol$^{-1}$ higher Au$_8^{I-}$ are shown in Figure 4 as well (see also Table 5). The HOMO-LUMO gaps of the open-shell cluster Au$_8^{I-}$ amounts to 1.01 eV (BP86/A) (cf. B3LYP: 1.69 [1.73; 1.68] eV) (spin-$\uparrow$) and 1.31 eV (BP86/A) (cf. B3LYP: 2.51 [2.53; 2.44] eV) (spin-$\downarrow$), respectively. The computed adiabatic electron affinity of Au$_8^{I-}$ is equal to 2.78 [2.74; 2.72] eV in agreement with the experimental magnitude of 2.764 $\pm$ 0.050 eV [33] and 2.73(2) eV [30] and the theoretical one of 2.88 eV [30]. The VDE of Au$_8^{I-}$ amounts to 2.88 [2.82; 2.82] eV which is also close to VDE$^{expt}$(Au$_8^-$) = 2.79 [33] and 2.79(2) eV [30] and VDE$^{theor}$(Au$_8^-$) = 3.0 eV (obtained with the BP86 and the RECP C augmented by additional polarization and diffuse functions optimized for the neutral Au$_3$ cluster [13]) and 2.94 eV [30]. Note that the 3D anion Au$_8^{III-}$ is stable under attaching e$^-$, that is, the dianion Au$_8^{III 2-}$, which becomes 2D and whose structure is displayed in Figure 4 (see also Table 5), is energetically lower than Au$_8^{III-}$ by 14.5 [11.8; 12.2] kcal$\cdot$mol$^{-1}$ and therefore exists as a long-lived metastable dianion.
\vspace{1cm}

\centerline{\bf {VII. The Low-Energy Section of the PES of Au$_9$}}
\vspace{0.25cm}

As we have seen in the previous Sections, the planar triangular motif works only for the neutral and anionic ground-state clusters Au$_{6-8}^{I}$ and not for the cationic ones. Does it mean that it is applicable for all larger clusters or is there some threshold size of gold clusters for which this structural rule is no longer valid? 

We show in this Section that $n = 9$ is in fact such a threshold. The structures of the low-energy nine gold atom clusters, displayed in Figure 5 (see also Table 6), supports this conjecture: three of them, a bicapped hexagonal ring Au$_9^{I}$, Au$_9^{III}$, and Au$_9^{IV}$ are 2D whereas Au$_9^{II}$ is 3D. The clusters Au$_9^{I}$, Au$_9^{II}$ and Au$_9^{IV}$ are novel and found in the present work for the first time (cf. the clusters in Refs. [4, 10]; see notes [53, 54]). The reported four structures of Au$_9$ can be considered as nearly isoenergetic due to the definition of the accepted of the DF/RECP methods for the gold clusters (see Ref. [10] and Section II). 

Comparing the neutral PESs of 6 - 9 gold atoms, we conclude that the three planar low-energy structures Au$_9^{I}$, Au$_9^{III}$ and Au$_9^{IV}$ originates from Au$_8^{I}$, Au$_7^{I}$ or Au$_6^{I}$ by adding one, two or three planar triangles, respectively. On the other hand, the almost isoenergetic 3D structure Au$_9^{II}$ is directly obtained from Au$_8^{I}$ by adding (capping) one gold atom to its central square and forming thus four new Au-Au bonds, instead of two as demanded by the planar triangular motif rule. This means that $n = 9$ is actually the minimal number of gold atoms for which the above rule is no longer fully valid. Therefore, the neutral PES of nine gold atoms markedly demonstrates the 2D - 3D coexistence on its low energy section.
\vspace{1cm}

\centerline{\bf {VIII. Summary and Outlook}}
\vspace{0.25cm}

A variety of novel low-energy 2D and 3D structures have been determined in the present work on the neutral, cationic and anionic PESs of small gold clusters Au$_{5 \leq n \leq 8}$ and on the neutral PES of Au$_9$. They provide new insights on the 2D - 3D transition in gold clusters, particularly on their well-known odd- vs. even-sized patterns. The neutral PES of Au$_5$ shows the presence of two new planar forms, Au$_5^{III}$ and Au$_5^{IV}$, filling thus the gap of $\approx$ 21 kcal$\cdot$mol$^{-1}$ between the 2D ground-state W-form and the 3D pyramid discussed in Ref. [10]. On the PES of Au$_7$, this gap is significantly reduced to $\approx$ 6 kcal$\cdot$mol$^{-1}$ due to the existence of the novel 3D cluster Au$_7^{II}$. For $n = 9$, the gap is essentially closed: the novel 3D structure Au$_9^{II}$ coexists on almost equal energetical footing with three 2D structures, two of which, including the slightly most stable one, Au$_9^{I}$, are new as well.  

As far as the 2D - 3D coexistence is concerned, the small even-numbered closed-shell clusters ($n \leq 10$) behave quite differently from the small neutral odd-numbered open-shell ones. This is particularly true for the PES of Au$_8$ where the lowest-energy 3D cluster Au$_8^{III}$ is separated from the 2D ground-state one by ca. 13 kcal$\cdot$mol$^{-1}$. We suggest that this is due to the stronger $s - d$ hybridization in Au$_{2n}$ than in Au$_{2n+1}$, at least for $2 \leq n \leq 4$.  Another feature related to the 2D - 3D transition in small odd-numbered clusters of gold emerges from the present work. This is the behavior of the gap E$_{HL}$. The spin-up gap E$_{HL \uparrow}$ of the ground-stable 2D clusters narrows with increasing $n$ from 1.57 eV (Au$_5^{I}$) to 1.32 (Au$_7^{I}$) and 0.90 eV (Au$_9^{I}$; the E$_{HL \uparrow}$ of Au$_9^{IV}$ = 0.22 eV). On the contrary, for $n$ = 7 and 9, the gap of the lowest-energy 3D clusters Au$_7^{II}$ and Au$_9^{II}$ widens from 0.92 to 1.32 eV. Note that E$_{HL \uparrow}$ of the 3D cluster Au$_7^{II}$ is smaller by 0.4 eV than that of the 2D one Au$_7^{I}$. This implies that the former has a higher catalytic activity despite its smaller binding energy and therefore deserves to be experimentally tested. The spin imbalance of the HOMO-LUMO gap, E$_{HL \uparrow}$ - E$_{HL \downarrow}$, decreases from 0.57 eV at $n = 5$ to 0.28 eV at $n = 7$ and 0.14 eV at $n = 9$ (Au$_9^I$).

Novel neutral clusters also lead to interesting features on the charged PESs. One example: the neutral cluster Au$_5^{III}$ gives rise on the anionic PES to the anion Au$_5^{III-}$ which has been found to be the most stable one, in contrast with previous experimental and theoretical works reporting that the ground anionic state is the W-shape structure Au$_5^{I-}$. This discrepancy can be explained by the presence of a pair of similar transition states, one between the ground-state W-form Au$_5^{I}$ and higher-energy cluster Au$_5^{III}$ on the neutral PES and another between Au$_5^{I-}$ and Au$_5^{III-}$ on the anionic PES. These transition states govern the breaking of two Au-Au bonds and are therefore characterized by rather high barriers (of at least 15 kcal$\cdot$mol$^{-1}$ on the neutral PES) which results in a lifetime of Au$_5^{I-}$ long enough to be experimentally observed. On the other hand, the Au$_5^{III-}$ isomer, though less reactive than Au$_5^{I-}$ due to a larger E$_{HL}$, would be interesting to detect as well for another two reasons. First, as demonstrated in Section III, it gives rise to the long-lived metastable 5-gold dianion whose structural pattern is still present in the 8-gold dianion Au$_8^{III2-}$. Second, the bonding of a single gold atom to the atom Au$_3$ of Au$_5^{III-}$ results in the formation of the anionic cluster Au$_6^{II-}$ which lies only 3 - 5 kcal$\cdot$mol$^{-1}$ higher the ground-state anion Au$_6^{I-}$. Both of them are built, by analogy with the triangular motif used for neutrals, via the bonding of Au to the respective 5-gold anions. 

As established above, for neutral clusters, the 2D - 3D coexistence takes place only for the odd-sized open-shell gold cluster Au$_9$. However, it already develops for the cationic odd-sized closed shell clusters Au$_5^+$ and Au$_7^+$, although the ways it does for these two $n$ are essentially different. For $n = 5$, the 2D - 3D transition emerges from the existence of two almost isoenergetic rotamers: the 2D Au$_5^{I+}$ and the 3D Au$_5^{I+\prime}$ which lacks its neutral 3D parent. On the other hand, the case $n$ = 7 is directly related to the ionization of the ground-state neutral 2D cluster Au$_7^{I}$. As noted in Section V, the ionization weakens the $s - d$ hybridization and thus makes the 2D triangular motif rule invalid. We suggest that this process occurs via the intermediate planar transition-state linker Au$_7^{I+ts}$ which relaxes into the 3D cluster Au$_7^{II+}$. In contrast to the 3D cation Au$_5^{I+\prime}$, Au$_7^{II+}$ possesses its 3D parent image Au$_7^{II}$ on the neutral PES which however, as already discussed above, is less stable than Au$_7^{I}$. Interestingly, the 3D Zn-analogue of Au$_7^{II+}$, where the Zn atom replaces Au$_4$ (or Au$_5$, see Figure 3), is only 3.2 kcal$\cdot$mol$^{-1}$ (B3LYP/A) higher the most planar stable cation Au$_6$Zn$^+$ reported in Ref. [20a] (the cluster [6a+] therein), that is, it falls within the DF/RECP error margin. 

In addition, the present computational study clearly demonstrates that another cationic cluster Au$_7^{III+}$ (see Figure 3), originating from the neutral planar honeycomb structure Au$_7^{III}$, is also 3D (at the B3LYP/A and B3LYP/B computational levels). As reported in Section V, it is almost isoenergetic to Au$_7^{II+}$ within the DF/RECP error margin. Moreover, the cation Au$_7^{IV+}$, separated from Au$_7^{II+}$ by about 9.6 kcal$\cdot$mol$^{-1}$ (B3LYP/A), is 3D too (see Figure 3). Altogether, this means that in the low-energy section of the PES of Au$_7^{+}$ , 3D structures are favored over the 2D ones up to energies of ca. 15 kcal$\cdot$mol$^{-1}$ (see the neutrals in Ref. [10]). In summary, the low-energy portion (within of ca. 15 kcal$\cdot$mol$^{-1}$ from the bottom) of the cationic PES of seven gold atoms is characterized by a complete absence of 2D clusters. Strickly speaking, the concept of the 2D - 3D coexistence might then be only valid beyond these energies.

The 2D - 3D coexistence in even-sized open-shell cationic clusters of gold is quite different from that of the odd-sized closed-shell cationic clusters and that of the even-sized closed-shell neutrals as well. As an example, the novel 3D cation Au$_6^{IV+}$ lies only 5 - 7 kcal$\cdot$mol$^{-1}$ higher than the almost degenerate ground-state cations Au$_6^{I+}$ and Au$_6^{III+}$. At $n$ = 8, the energy offset between the 2D ground-state cluster Au$_8^{II+}$ and 3D one Au$_8^{IV+}$ reduces to $\approx$ 1 kcal$\cdot$mol$^{-1}$ which falls within the DF/RECP computational error. One may therefore anticipate that the even $n$ = 10 might show the clearcut 2D - 3D coexistence at the bottom of the PES.

The final comment is about the anionic PESs of gold clusters. It has been demonstrated in the present work that the studied anions hardly manifest the 2D - 3D coexistence. This conclusion is in agreement with the early findings by Landman and co-workers [4, 30]. The most stable 3D cluster which has been reported in this work for the first time is Au$_8^{III-}$. However, it is strongly (by about 21 - 23 kcal$\cdot$mol$^{-1}$) separated from the 2D ground-state cluster Au$_8^{I-}$. 
\vspace{1cm}

\centerline{\bf {Acknowledgments}}
\vspace{0.25cm}

This work was supported by the Region Wallonne (RW. 115012). The computational facilities were provided by NIC (University of Liege) and by F.R.F.C. 9.4545.03 (FNRS, Belgium). The authors thank the reviewer for valuable comments and suggestions.

\pagebreak

\pagebreak

\centerline \normalsize{\bf {Figure Legends}}
\vspace{1cm}

Figure 1. The low-energy sections of the neutral, cationic and anionic PESs of Au$_5$ and the structure of the long-lived dianion Au$_5^{III 2-}$.  
\vspace{1cm}

Figure 2. The low-energy portions of the neutral, cationic and anionic PESs of six gold atoms.
\vspace{1cm}

Figure 3. The low-energy portions of the neutral, cationic and anionic PESs of seven gold atoms.
\vspace{1cm}

Figure 4. The low-energy portions of the neutral, cationic and anionic PESs of Au$_8$ and the structure of the long-lived dianion Au$_8^{III 2-}$.  
\vspace{1cm}

Figure 5. The low-energy portion of the neutral PES of nine atoms of gold. The bond lengths of the displayed clusters are listed in Table 6.
\vspace{1cm}

\begin{landscape}
\begin{table}
\caption{The bond lengths (in {\AA}) of the low-energy neutral and charged clusters of five gold atoms.}
\vspace{0.25cm}
\tiny{
\begin{tabular}{|l|cccccc|}
\hline
&&&&&&\\
&B3LYP/A&B3LYP/B&B3LYP/C&BP86/A&PW91/A&Data\\
&&&&&&\\
\hline\hline
&&&&&&\\
{\bf Au$_5^{I} ({}^2A_1\, C_{2v})$}&&&&&&\\
r(Au$_1$-Au$_2$)&2.872&2.831&2.863&2.837&2.829&2.78 [4]; 2.81 [7]; 2.704 [8]\\
r(Au$_1$-Au$_3$)&2.698&2.678&2.702&2.667&2.666&2.64 [4], 2.72 [7], 2.566 [8]\\
r(Au$_2$-Au$_3$)&2.750&2.722&2.752&2.723&2.721&2.75 [6]; 2.614 [8]\\
r(Au$_2$-Au$_5$)&2.777&2.740&2.777&2.734&2.730&2.79 [6]; 2.609 [8]\\
&&&&&&\\
{\bf Au$_5^{I-} ({}^1A_1\, C_{2v})$}&&&&&&\\
r(Au$_1$-Au$_2$)&3.001&2.906&2.959&&&2.82 [4]; 2.725 [8]\\
r(Au$_1$-Au$_3$)&2.679&2.666&2.685&&&2.65 [4]; 2.566 [8]\\
r(Au$_2$-Au$_3$)&2.881&2.831&2.877&&&2.672 [8]\\
r(Au$_2$-Au$_5$)&2.676&2.831&2.877&&&2.582 [8]\\
&&&&&&\\
{\bf Au$_5^{I+} ({}^1B_{2u}\, D_{2h})$}&&&&&&\\
r(Au$_1$-Au$_2$)&2.765&2.736&2.767&&&2.684 [14]\\
r(Au$_2$-Au$_5$)&2.643&2.628&2.650&&&2.588 [14]\\
&&&&&&\\
{\bf Au$_5^{I+ \prime}$}&&&&&&\\
r(Au$_1$-Au$_2$)&2.768&2.629&&&&\\
r(Au$_2$-Au$_3$)&2.645&2.629&&&&\\
&&&&&&\\
{\bf Au$_5^{II} ({}^2B_{2u}\, D_{2h})$}&&&&&&\\
r(Au$_1$-Au$_2$)&2.714&2.689&2.716&2.684&&\\
r(Au$_2$-Au$_5$)&2.751&2.722&2.751&2.723&&2.609 [8]\\
&&&&&&\\
{\bf Au$_5^{III}$}&&&&&&\\
r(Au$_1$-Au$_2$)&2.619&2.603&2.631&2.594&&\\
r(Au$_1$-Au$_3$)&2.731&2.707&2.729&2.684&&\\
r(Au$_1$-Au$_4$)&2.793&2.742&2.793&2.742&&\\
r(Au$_4$-Au$_5$)&2.619&2.601&2.629&2.586&&\\
&&&&&&\\
{\bf Au$_5^{III-}$}&&&&&&\\
r(Au$_1$-Au$_2$)&2.628&2.613&2.635&&&\\
r(Au$_1$-Au$_3$)&2.719&2.697&2.718&&&\\
r(Au$_1$-Au$_4$)&2.922&2.697&2.718&&&\\
&&&&&&\\
{\bf Au$_5^{III2-}$}&&&&&&\\
r(Au$_1$-Au$_2$)&2.740&2.707&2.728&&&\\
r(Au$_1$-Au$_3$)&2.867&2.818&2.848&&&\\
r(Au$_1$-Au$_4$)&2.734&2.715&2.746&&&\\
&&&&&&\\
{\bf Au$_5^{IV}$}&&&&&&\\
r(Au$_1$-Au$_3$)&2.872&2.814&2.839&&&\\
r(Au$_1$-Au$_4$)&2.643&2.640&2.645&&&\\
r(Au$_2$-Au$_5$)&2.650&2.626&2.662&&&\\
r(Au$_3$-Au$_4$)&2.744&2.701&2.769&&&\\
r(Au$_3$-Au$_5$)&2.643&2.617&2.657&&&\\
&&&&&&\\
\hline
\end{tabular}
}
\end{table}
\end{landscape}
\begin{table}
\caption{The relative energies (in kcal$\cdot$mol$^{-1}$) $\Delta$E and ZPVE-corrected ones $\Delta$E$_{ZPVE}$ (in parentheses) and the HOMO-LUMO gaps E$_{HL}$ (in eV) of the neutral, cationic and anionic gold clusters Au$_{5 \leq n \leq 9}$. Notice that E$_{HL \uparrow}$(Au$_9^{IV}$) = 0.46 eV [29].}
\vspace{0.25cm}
\tiny{
\begin{tabular}{|l|cccc|cccc|}
\hline
&&&&&&&&\\
&\multicolumn{4}{c|}{$\Delta$E ($\Delta$E$_{ZPVE}$)}&\multicolumn{4}{c|}{E$_{HL \uparrow}$ E$_{HL \downarrow}$}\\
&B3LYP/A&B3LYP/B&B3LYP/C&BP86/A&B3LYP/A&B3LYP/B&B3LYP/C&BP86/A\\
&&&&&&&&\\
\hline\hline
&&&&&&&&\\
{\bf Au$_5$}&&&&&&&&\\ 
Au$_5^I$&0.0&0.0&0.0&0.0&2.49 2.17&2.46 2.17&2.47 2.17&1.58 1.01\\
Au$_5^{II}$&7.28 (7.27)&8.17 (8.15)&7.99 (7.97)&8.04 (8.03)&1.44 2.84&1.50 2.73&1.43 2.98&0.77 1.31\\
Au$_5^{III}$&13.88 (13.80)&15.93 (15.83)&14.92 (14.81)&15.16 (15.15)&2.68 1.38&2.65 1.42&2.62 1.34&1.83 0.50\\
Au$_5^{IV}$&18.56 (18.40)&20.51 (20.34)&19.37 (19.18)&&1.16 2.36&1.12 2.46&1.24 2.10&\\
&&&&&&&&\\
{\bf Au$_5^-$}&&&&&&&&\\
Au$_5^{III-}$&0.0&0.0&0.0&0.0&2.88&2.39&2.41&\\
Au$_5^{I-}$&7.21 (7.19)&5.82 (5.84)&6.30 (6.30)&&1.21&0.93&1.09&\\
&&&&&&&&\\
{\bf Au$_6$}&&&&&&&&\\
Au$_6^{I}$&&&&&3.48&3.35&3.45&2.23\\
&&&&&&&&\\
{\bf Au$_6^-$}&&&&&&&&\\
Au$_6^{I-}$&0.0&0.0&0.0&0.0&1.11 3.30&1.15 3.27&1.10 3.27&0.40 2.09\\
Au$_6^{II-}$&2.52 (2.46)&5.78 (5.66)&4.36 (4.25)&&2.80 1.84&2.82 1.86&2.84 1.74&1.91 0.86\\
&&&&&&&&\\
{\bf Au$_7$}&&&&&&&&\\ 
Au$_7^{I}$&0.0&0.0&0.0&0.0&2.24 2.25&2.19 2.20&2.20 2.27&1.32 1.04\\
Au$_7^{II}$&5.56 (5.51)&6.46 (6.41)&5.73 (5.43)&5.08 (5.02)&1.83 3.04&1.74 3.11&1.78 3.10&0.92 1.79\\
Au$_7^{III}$&8.99 (8.82)&7.69 (7.52)&8.15 (7.98)&8.07 (7.87)&&&&\\
&&&&&&&&\\
{\bf Au$_7^-$}&&&&&&&&\\ 
Au$_7^{I-}$&0.0&0.0&0.0&0.0&1.95&1.94&1.89&1.18\\
Au$_7^{II-}$&4.56 (4.45)&7.87 (7.70)&6.01 (5.87)&2.85&2.77&2.79&1.72&\\
Au$_7^{III-}$&-2.51 (-2.50)&-1.71 (-1.73)&-2.44 (-2.44)&-0.87 (-0.87)&2.17&2.17&2.12&1.35\\
&&&&&&&&\\
{\bf Au$_8$}&&&&&&&&\\ 
Au$_8^{I}$&0.0&0.0&0.0&0.0&2.77&2.72&2.67&1.74\\
Au$_8^{II}$&12.53 (12.44)&10.77 (10.72)&11.01 (10.94)&9.67 (9.62)&1.76&1.71&1.72&0.87\\
&&&&&&&&\\
{\bf Au$_9$}&&&&&&&&\\ 
Au$_9^{I}$&0.0&0.0&0.0&0.0&1.80 1.66&1.74 1.65&1.77 1.62&0.90 0.76\\
Au$_9^{II}$&1.98 (1.99)&4.55 (4.53)&4.33 (4.30)&1.72 (1.71)&2.21 1.92&2.19 1.91&2.19 1.85&1.32 1.08\\
Au$_9^{III}$&3.22 (3.21)&3.24 (3.03)&3.03 (3.02)&2.56 (2.56)&2.12 1.03&2.05 1.04&2.07 1.00&1.22 0.19\\
Au$_9^{IV}$&4.36 (4.26)&4.56 (4.52)&3.85 (3.05)&3.55 (3.48)&0.91 2.39&0.92 2.33&0.89 2.40&0.22 1.35\\
&&&&&&&&\\
\hline
\end{tabular}
}
\end{table}

\begin{table}
\caption{The bond lengths (in {\AA}) of Au$_6^{I}$, Au$_6^{I+}$, Au$_6^{I-}$, Au$_6^{II-}$, Au$_6^{III+}$, and Au$_6^{IV+}$.}
\vspace{0.25cm}
\tiny{
\begin{tabular}{|l|ccccc|}
\hline
&&&&&\\
&B3LYP/A&B3LYP/B&B3LYP/C&BP86/A&Data\\
&&&&&\\
\hline\hline
&&&&&\\
{\bf Au$_6^{I} ({}^1A_{1}^\prime\, D_{3h})$}&&&&&\\
r(Au$_1$-Au$_2$)&2.900&2.856&2.890&2.856&2.81 [4]; 2.84 [7] \\
r(Au$_1$-Au$_3$)&2.900&2.856&2.890&2.855&2.81 [4]; 2.84 [7] \\
r(Au$_1$-Au$_4$)&2.704&2.683&2.706&2.677&2.66 [4]; 2.72 [7] \\
r(Au$_1$-Au$_5$)&2.704&2.683&2.706&2.677&2.66 [4]; 2.72 [7]\\
r(Au$_3$-Au$_4$)&2.704&2.683&2.706&2.677&2.66 [4]; 2.72 [7] \\
&&&&&\\
{\bf Au$_6^{I+}$}&&&&&\\
r(Au$_1$-Au$_2$)&2.997&2.930&3.076&&2.915 [14] \\
r(Au$_1$-Au$_3$)&2.732&2.708&2.804&&2.752 [14] \\
r(Au$_1$-Au$_4$)&2.728&2.688&2.694&&2.623 [14] \\
r(Au$_1$-Au$_5$)&2.700&2.674&2.733&&\\
r(Au$_3$-Au$_4$)&2.704&2.679&2.727&&\\
&&&&&\\
{\bf Au$_6^{I-} ({}^2A_{1}^\prime\, D_{3h})$}&&&&&\\
r(Au$_1$-Au$_2$)&2.810&2.781&2.808&2.784&2.76 [4] \\
r(Au$_1$-Au$_3$)&2.810&2.781&2.808&2.784&2.76 [4] \\
r(Au$_1$-Au$_4$)&2.781&2.746&2.778&2.743&2.71 [4] \\
r(Au$_1$-Au$_5$)&2.781&2.746&2.778&2.743&2.71 [4] \\
r(Au$_3$-Au$_4$)&2.781&2.746&2.778&2.743&2.71 [4] \\
&&&&&\\
{\bf Au$_6^{II-} (D_{3h})$}&&&&&\\
r(Au$_1$-Au$_2$)&2.764&2.728&2.765&2.715&\\
r(Au$_1$-Au$_4$)&2.645&2.623&2.654&2.614&\\
&&&&&\\
{\bf Au$_6^{III+}$}&&&&&\\
r(Au$_1$-Au$_2$)&2.700&2.680&2.704&&\\
r(Au$_1$-Au$_3$)&2.736&2.703&2.739&&\\
r(Au$_2$-Au$_3$)&2.803&2.778&2.802&&\\
r(Au$_2$-Au$_4$)&2.824&2.777&2.820&&\\
r(Au$_3$-Au$_4$)&2.868&2.821&2.865&&\\
&&&&&\\
{\bf Au$_6^{IV+}$}&&&&&\\
r(Au$_1$-Au$_2$)&2.720&2.720&2.729&&\\
r(Au$_1$-Au$_3$)&2.744&2.726&2.753&&\\
r(Au$_1$-Au$_4$)&2.948&2.888&2.946&&\\
r(Au$_1$-Au$_5$)&2.999&2.892&2.957&&\\
r(Au$_2$-Au$_3$)&3.079&2.948&2.997&&\\
r(Au$_2$-Au$_5$)&2.822&2.783&2.831&&\\
r(Au$_3$-Au$_4$)&2.783&2.773&2.794&&\\
r(Au$_4$-Au$_6$)&2.712&2.695&2.715&&\\
r(Au$_5$-Au$_6$)&2.722&2.699&2.726&&\\
&&&&&\\
\hline
\end{tabular}
}
\end{table}

\begin{table}
\caption{The bond lengths (in {\AA}) of the clusters Au$_7^{I-III}$ and their charged cousins. The superscript a indicates the bond length equal to 2.714 {\AA} in Ref. [4].}
\vspace{0.25cm}
\tiny{
\begin{tabular}{|l|cccc|l|cccc|}
\hline
&&&&&&&&&\\
&B3LYP/A&B3LYP/B&B3LYP/C&BP86/A&&B3LYP/A&B3LYP/B&B3LYP/C&BP86/A\\
&&&&&&&&&\\
\hline\hline
&&&&&&&&&\\
{\bf Au$_7^{I} ({}^2A^\prime\, C_{s})$}&&&&&{\bf Au$_7^{I+ts}$}&&&&\\
r(Au$_1$-Au$_2$)&2.769&2.734&2.763&2.745&r(Au$_1$-Au$_2$)&2.710&&&\\
r(Au$_1$-Au$_4$)&2.756&2.725&2.762&2.711&r(Au$_1$-Au$_4$)&2.713&&&\\
r(Au$_2$-Au$_3$)&2.717&2.693&2.718&2.683&r(Au$_2$-Au$_3$)&2.810&&&\\
r(Au$_2$-Au$_4$)&2.780&2.753&2.779&2.753&r(Au$_2$-Au$_4$)&2.858&&&\\
r(Au$_3$-Au$_4$)&3.102&3.003&3.065&2.980&r(Au$_3$-Au$_4$)&2.903&&&\\
r(Au$_3$-Au$_6$)&2.868&2.829&2.860&2.843&r(Au$_3$-Au$_6$)&2.918&&&\\
r(Au$_3$-Au$_7$)&2.724&2.702&2.728&2.698&r(Au$_3$-Au$_7$)&2.667&&&\\
r(Au$_4$-Au$_5$)&2.712&2.689&2.715&2.687&r(Au$_4$-Au$_5$)&2.701&&&\\
r(Au$_4$-Au$_6$)&2.845&2.802&2.840&2.794&r(Au$_4$-Au$_6$)&2.782&&&\\
r(Au$_5$-Au$_6$)&2.695&2.674&2.696&2.669&r(Au$_5$-Au$_6$)&2.700&&&\\
r(Au$_6$-Au$_7$)&2.713&2.688&2.713&2.683&r(Au$_6$-Au$_7$)&2.754&&&\\
&&&&&&&&&\\
{\bf Au$_7^{II}$}&&&&&{\bf Au$_7^{II+}$}&&&&\\
r(Au$_1$-Au$_2$)&2.706&2.685&2.708&2.679&r(Au$_1$-Au$_2$)&2.725&2.699&2.726&\\
r(Au$_1$-Au$_5$)&2.701&2.678&2.703&2.674&r(Au$_1$-Au$_5$)&2.671&2.654&2.675&\\
r(Au$_2$-Au$_3$)&2.706&2.685&2.708&2.679&r(Au$_2$-Au$_3$)&2.725&2.699&2.726&\\
r(Au$_2$-Au$_4$)&2.899&2.855&2.890&2.856&r(Au$_2$-Au$_4$)&2.876&2.841&2.871&\\
r(Au$_2$-Au$_5$)&2.899&2.855&2.890&2.856&r(Au$_2$-Au$_5$)&2.871&2.841&2.871&\\
r(Au$_3$-Au$_4$)&2.701&2.678&2.703&2.674&r(Au$_3$-Au$_4$)&2.671&2.654&2.675&\\
r(Au$_4$-Au$_5$)&2.966&2.917&2.948&2.914&r(Au$_4$-Au$_5$)&2.995&2.934&2.979&\\
r(Au$_4$-Au$_6$)&2.756&2.786&2.816&2.777&r(Au$_4$-Au$_6$)&2.861&2.822&2.860&\\
r(Au$_4$-Au$_7$)&2.901&2.786&2.816&2.777&r(Au$_4$-Au$_7$)&2.861&2.822&2.860&\\
r(Au$_5$-Au$_6$)&2.901&2.786&2.816&2.777&r(Au$_5$-Au$_6$)&2.861&2.822&2.860&\\
r(Au$_5$-Au$_7$)&2.756&2.786&2.816&2.777&r(Au$_5$-Au$_7$)&2.861&2.822&2.860&\\
r(Au$_6$-Au$_7$)&2.916&2.869&2.923&2.885&r(Au$_6$-Au$_7$)&2.861&2.670&2.685&\\
&&&&&&&&&\\
{\bf Au$_7^{III}$}&&&&&{\bf Au$_7^{IV}$}&&&&\\
r(Au$_1$-Au$_2$)&2.840&2.725&2.753&2.732&r(Au$_1$-Au$_2$)&2.881&&&\\
r(Au$_1$-Au$_4$)&2.750&2.732&2.805&2.717&r(Au$_1$-Au$_3$)&2.795&&&\\
r(Au$_3$-Au$_7$)&2.910&2.791&2.766&2.854&r(Au$_2$-Au$_3$)&2.818&&&\\
r(Au$_4$-Au$_5$)&2.842&2.860&2.860&2.802&r(Au$_2$-Au$_5$)&2.697&&&\\
r(Au$_5$-Au$_6$)&2.756&2.810&2.905&2.861&r(Au$_2$-Au$_6$)&2.797&&&\\
r(Au$_6$-Au$_7$)&2.755&2.721&2.751&2.733&r(Au$_3$-Au$_6$)&2.695&&&\\
&&&&&&&&&\\
{\bf Au$_7^{III+}$}&&&&&{\bf Au$_7^{IVts}$}&&&&\\
r(Au$_1$-Au$_2)^a$&2.807&2.773&2.790&2.777&r(Au$_1$-Au$_2$)&2.751&&&\\
r(Au$_1$-Au$_4$)&2.779&2.747&2.790&2.750&r(Au$_2$-Au$_6$)&2.744&&&\\
&&&&&&&&&\\
{\bf Au$_7^{I-} ({}^1A_{1}\, C_{2v})$}&&&&&{\bf Au$_7^{III-}$}&&&&\\
r(Au$_1$-Au$_2$)&2.837&2.788&2.830&2.783&r(Au$_1$-Au$_2$)&2.655&2.635&2.662&2.625\\
r(Au$_1$-Au$_7$)&2.648&2.640&2.655&2.634&r(Au$_1$-Au$_3$)&2.757&2.725&2.758&2.718\\
r(Au$_2$-Au$_3$)&2.730&2.712&2.733&2.712&r(Au$_3$-Au$_4$)&2.766&2.752&2.765&2.763\\
r(Au$_3$-Au$_4$)&2.719&2.697&2.719&2.694&r(Au$_3$-Au$_5$)&2.789&2.752&2.785&2.748\\
r(Au$_3$-Au$_5$)&2.874&2.807&2.862&2.802&r(Au$_5$-Au$_6$)&2.692&2.677&2.696&2.672\\
r(Au$_5$-Au$_7$)&2.797&2.753&2.791&2.752&&&&&\\
&&&&&&&&&\\
{\bf Au$_7^{II-}$}&&&&&&&&&\\
r(Au$_1$-Au$_4$)&2.692&2.675&2.694&2.670&&&&&\\
r(Au$_1$-Au$_5$)&2.882&2.822&2.868&2.810&&&&&\\
r(Au$_2$-Au$_3$)&2.727&2.700&2.729&2.695&&&&&\\
r(Au$_2$-Au$_6$)&2.622&2.608&2.629&2.597&&&&&\\
&&&&&&&&&\\
\hline
\end{tabular}
}
\end{table}

\begin{table}
\caption{The bond lengths (in {\AA}) of the neutral, cationic and anionic clusters of Au$_8$.}
\vspace{0.25cm}
\tiny{
\begin{tabular}{|l|cccc|l|cccc|}
\hline
&&&&&&&&&\\
&B3LYP/A&B3LYP/B&B3LYP/C&BP86/A&&B3LYP/A&B3LYP/B&B3LYP/C&BP86/A\\
&&&&&&&&&\\
\hline\hline
&&&&&&&&&\\
{\bf Au$_8^{I} ({}^1A_{1g}\, D_{4h})$}&&&&&{\bf Au$_8^{III}$}&&&&\\
r(Au$_1$-Au$_2$)&2.697&2.675&2.698&2.670&r(Au$_1$-Au$_2$)&2.817&2.778&2.812&2.775\\
r(Au$_1$-Au$_5$)&2.697&2.675&2.698&2.670&r(Au$_1$-Au$_3$)&2.658&2.643&2.664&2.637\\
r(Au$_2$-Au$_4$)&2.870&2.826&2.860&2.819&r(Au$_2$-Au$_7$)&2.978&2.938&2.968&2.934\\
r(Au$_4$-Au$_5$)&4.066&4.003&4.045&3.980&r(Au$_2$-Au$_8$)&2.817&2.778&2.812&2.773\\
&&&&&r(Au$_4$-Au$_5$)&2.658&2.643&2.665&2.637\\
{\bf Au$_8^{I+} ({}^2A_{1g}\, D_{4h})$}&&&&&r(Au$_5$-Au$_6$)&2.784&2.758&2.784&2.753\\
r(Au$_1$-Au$_2$)&2.722&2.695&2.720&&{\bf Au$_8^{III-}$}&&&&\\
r(Au$_1$-Au$_5$)&2.704&2.679&2.711&&r(Au$_1$-Au$_2$)&2.774&2.747&2.776&\\
r(Au$_2$-Au$_4$)&2.844&2.805&2.840&&r(Au$_2$-Au$_4$)&2.935&2.887&2.929&\\
r(Au$_4$-Au$_5$)&3.004&2.919&2.990&&r(Au$_3$-Au$_6$)&2.920&2.887&2.913&\\
&&&&&r(Au$_3$-Au$_8$)&2.763&2.727&2.757&\\
{\bf Au$_8^{I-} ({}^2A_{1g} D_{4h})$}&&&&&r(Au$_4$-Au$_6$)&3.040&3.000&3.038&\\
r(Au$_1$-Au$_2$)&2.754&2.722&2.752&2.721&r(Au$_4$-Au$_7$)&2.707&2.690&2.713&\\
r(Au$_1$-Au$_5$)&2.754&2.722&2.752&2.721&r(Au$_6$-Au$_7$)&2.796&2.758&2.791&\\
r(Au$_2$-Au$_4$)&2.782&2.752&2.780&2.751&r(Au$_6$-Au$_8$)&2.805&2.779&2.807&\\
r(Au$_4$-Au$_5$)&3.940&3.893&3.931&3.895&{\bf Au$_8^{III2-}$}&&&&\\
&&&&&&&&&\\
{\bf Au$_8^{II}$}&&&&&r(Au$_1$-Au$_2$)&2.732&2.705&2.736&\\
r(Au$_1$-Au$_2$)&2.946&2.866&2.924&2.866&r(Au$_2$-Au$_3$)&2.715&2.697&2.716&\\
r(Au$_2$-Au$_3$)&2.771&2.743&2.766&2.764&r(Au$_2$-Au$_4$)&2.840&2.800&2.844&\\
r(Au$_3$-Au$_4$)&2.809&2.766&2.805&2.758&r(Au$_3$-Au$_4$)&2.777&2.745&2.773&\\
r(Au$_3$-Au$_6$)&2.877&2.827&2.866&2.816&r(Au$_3$-Au$_5$)&2.779&2.741&2.773&\\
r(Au$_5$-Au$_6$)&2.733&2.714&2.736&2.706&r(Au$_3$-Au$_6$)&3.076&2.943&3.012&\\
r(Au$_6$-Au$_7$)&2.839&2.804&2.832&2.822&r(Au$_4$-Au$_6$)&3.378&3.275&3.287&\\
r(Au$_7$-Au$_8$)&2.772&2.705&2.729&2.700&r(Au$_4$-Au$_7$)&2.668&2.654&2.674&\\
&&&&&r(Au$_5$-Au$_6$)&2.733&2.715&2.740&\\
{\bf Au$_8^{II+}$}&&&&&r(Au$_6$-Au$_7$)&2.795&2.753&2.786&\\
r(Au$_1$-Au$_2$)&2.828&2.782&2.822&&r(Au$_7$-Au$_8$)&2.962&2.877&2.933&\\
r(Au$_2$-Au$_3$)&2.813&2.782&2.806&&{\bf Au$_8^{IV}$}&&&&\\
r(Au$_3$-Au$_4$)&2.789&2.756&2.787&&
r(Au$_1$-Au$_2$)&2.703&2.681&2.706&\\
r(Au$_3$-Au$_6$)&2.819&2.774&2.816&&r(Au$_2$-Au$_3$)&2.886&2.846&2.875&\\
r(Au$_5$-Au$_6$)&2.886&2.730&2.759&&r(Au$_2$-Au$_6$)&2.740&2.723&2.743&\\
r(Au$_6$-Au$_7$)&2.886&2.852&2.875&&r(Au$_3$-Au$_5$)&2.970&2.906&2.960&\\
r(Au$_7$-Au$_8$)&2.724&2.699&2.728&&r(Au$_4$-Au$_5$)&2.672&2.670&2.682&\\
&&&&&r(Au$_5$-Au$_8$)&3.033&2.958&3.011&\\
{\bf Au$_8^{II-}$}&&&&&r(Au$_7$-Au$_8$)&2.729&2.711&2.731&\\
r(Au$_1$-Au$_2$)&2.890&2.829&2.877&&{\bf Au$_8^{IV+}$}&&&&\\
r(Au$_2$-Au$_3$)&2.739&2.716&2.740&&r(Au$_1$-Au$_2$)&2.696&2.675&2.702&\\
r(Au$_3$-Au$_4$)&2.857&2.811&2.877&&r(Au$_2$-Au$_3$)&2.871&2.844&2.665&\\
r(Au$_3$-Au$_6$)&2.927&2.869&2.910&&r(Au$_2$-Au$_4$)&2.936&2.891&2.936&\\
r(Au$_5$-Au$_6$)&2.723&2.706&2.728&&r(Au$_2$-Au$_7$)&2.977&2.934&2.975&\\
r(Au$_6$-Au$_7$)&3.008&2.940&2.979&&r(Au$_3$-Au$_8$)&2.786&2.759&2.786&\\
r(Au$_7$-Au$_8$)&2.744&2.717&2.745&&r(Au$_4$-Au$_5$)&2.795&2.765&2.797&\\
{\bf Au$_8^{V}$}&&&&&r(Au$_4$-Au$_7$)&2.931&2.895&2.930&\\
r(Au$_1$-Au$_2$)&2.850&2.821&2.849&&r(Au$_5$-Au$_8$)&2.834&2.799&2.831&\\
r(Au$_4$-Au$_5$)&2.759&2.740&2.759&&r(Au$_7$-Au$_8$)&2.757&2.734&2.759&\\
r(Au$_5$-Au$_8$)&2.853&2.808&2.849&&&&&&\\
r(Au$_7$-Au$_8$)&2.721&2.707&2.725&&&&&&\\
&&&&&&&&&\\
\hline
\end{tabular}
}
\end{table}

\begin{table}
\caption{The bond lengths (in {\AA}) of the neutral clusters of nine gold atoms.}
\vspace{0.25cm}
\small{
\begin{tabular}{|l|cccc|}
\hline
&&&&\\
&B3LYP/A&B3LYP/B&B3LYP/C&BP86/A\\
&&&&\\
\hline\hline
&&&&\\
{\bf Au$_9^I ({}^2A_{1}\, C_{2v})$}&&&&\\ 
r(Au$_1$-Au$_2$)&2.719&2.698&2.726&2.694\\
r(Au$_1$-Au$_3$)&2.902&2.850&2.885&2.857\\
r(Au$_1$-Au$_4$)&3.161&3.033&3.092&2.991\\ 
r(Au$_3$-Au$_4$)&2.776&2.745&2.775&2.747\\
r(Au$_4$-Au$_5$)&2.769&2.753&2.774&2.764\\
r(Au$_4$-Au$_6$)&2.788&2.746&2.787&2.748\\
r(Au$_6$-Au$_7$)&2.868&2.810&2.851&2.805\\
&&&&\\
{\bf Au$_9^{II} ({}^2A_{1}\, C_{4v})$}&&&&\\
r(Au$_1$-Au$_2$)&2.912&2.856&2.909&2.851\\
r(Au$_3$-Au$_4$)&2.922&2.887&2.919&2.880\\
r(Au$_4$-Au$_6$)&2.710&2.688&2.711&2.684\\
&&&&\\
{\bf Au$_9^{III} ({}^2A_{1}\, D_{2h})$}&&&&\\
r(Au$_1$-Au$_2$)&2.761&2.733&2.758&2.737\\
r(Au$_1$-Au$_3$)&2.956&2.882&2.930&2.884\\
r(Au$_1$-Au$_4$)&2.779&2.751&2.782&2.753\\
r(Au$_3$-Au$_4$)&2.752&2.725&2.756&2.709\\
r(Au$_3$-Au$_9$)&2.670&2.658&2.677&2.650\\
r(Au$_4$-Au$_6$)&2.891&2.831&2.868&2.849\\
r(Au$_2$-Au$_5$)&2.706&2.687&2.711&2.676\\
r(Au$_5$-Au$_7$)&2.950&2.885&2.930&2.905\\
r(Au$_5$-Au$_8$)&2.802&2.759&2.797&2.755\\
&&&&\\
{\bf Au$_9^{IV}$}&&&&\\
r(Au$_1$-Au$_3$)&2.876&2.729&2.793&2.830\\
r(Au$_2$-Au$_3$)&2.745&2.729&2.746&2.703\\
r(Au$_1$-Au$_4$)&2.876&2.840&2.866&2.798\\
r(Au$_1$-Au$_6$)&2.844&2.755&2.834&2.796\\
r(Au$_1$-Au$_7$)&2.844&2.842&2.837&2.833\\
r(Au$_4$-Au$_5$)&2.729&2.753&2.726&2.674\\
r(Au$_4$-Au$_6$)&3.019&2.928&3.023&3.015\\
r(Au$_5$-Au$_6$)&2.679&2.645&2.686&2.675\\
&&&&\\
\hline
\end{tabular}
}
\end{table}

\begin{thebibliography}{99}
\bibitem{a31} Au$_5^{III-}$ is distinguished from the V-shaped chain, found in Ref. [13] and placed 0.13 eV higher Au$_5^{I-}$, due to the presence of the additional bond Au$_1$-Au$_4$ (see Figure 1).
\bibitem{a47} Note that the planar structure II reported to be lying 2.26 kcal$\cdot$mol$^{-1}$ higher relative to Au$_7^{I}$ on the PES of Au$_7$ in Ref. [10] or 0.1 kcal$\cdot$mol$^{-1}$ lower in Ref. [3] (see Figure 1 therein), has not been found in the present work at the B3LYP/A computational level. On the contrary, when taken as the initial geometry, it converges to Au$_7^{I}$ under the tight optimization.
\bibitem{a48} Note that the anionic structure 7B reported in Ref. [30] and resembling the structure of the neutral cluster Au$_7^{I}$, lying 0.12 kcal$\cdot$mol$^{-1}$ higher than Au$_7^{I-}$ on the PES of Au$_7^-$, has not been found in the present work at the B3LYP/A computational level. As discussed for note [47], when taken as the initial guess, it converges to Au$_7^{I-}$ under the tight optimization.
\bibitem{a49} This is a difference between the results for the ground-state structure on the PES of the neutral gold cluster of seven atoms, presented in Refs. [4, 10] and this work. Note that in Ref. [4], Au$_7^{III}$ was predicted to be the global minimum.
\bibitem{a50} In some sense, Au$_7^{III+}$ resembles the 2D structure of the cation Au$_7^+$ with the $D_{6h}$ symmetry obtained in Ref. [14]. However, the present B3LYP/A and B3LYP/B calculations show that Au$_7^{III+}$ is undoubtedly a  highly-symmetric 3D structure.
\bibitem{a51} It is suggested in Ref. [14] (p. 4099) that "... several three dimensional structures ... can be ruled out since the lowest is almost 0.5 eV higher in energy and the cross sections also disagree with measurement - the pentagonal bipyramid (the lowest 3D structure we found) for example, has a cross section of 61.7 {\AA}$^2$ ...". The experimental and computed cross  sections of Au$_7^{III+}$ are correspondingly equal to 65.8 and 65.9 {\AA}$^2$ [14]. It would be interesting to calculate the collisional cross section of Au$_7^{I+}$.
\bibitem{a53} Due to the absence of the structures Au$_9^{I}$ and Au$_9^{II}$ in Ref. [10], the most stable cluster was assigned to Au$_9^{III}$.
\bibitem{a54} All structures Au$_9^{I-IV}$ are found on the anionic PES in Ref. [13]. Therein, Au$_9^{III-}$ is the most stable one followed by Au$_9^{I-}$ with the energy difference of 4.8 kcal$\cdot$mol$^{-1}$. The next is a 3D structure Au$_9^{II-}$ separated from Au$_9^{III-}$ by 9.0 kcal$\cdot$mol$^{-1}$ and the last is a planar cluster Au$_9^{IV-}$ lying 17.8 kcal$\cdot$mol$^{-1}$ higher Au$_9^{III-}$.
\end{thebibliography}
\end{document}